\documentclass[preprint2]{aastex}

\usepackage{txfonts}
\usepackage{natbib}
\usepackage{graphicx}
\usepackage{times}
\bibpunct{(}{)}{;}{a}{}{,}
\newcommand{\Mnom}{\hbox{$M_\odot^{2010}$}}
\newcommand{\Rnom}{\hbox{$\mathcal{R}^{\mathrm N}_\odot$}}
\newcommand{\Lnom}{\hbox{$\mathcal{L}^{\mathrm N}_\odot$}}
\newcommand{\GMn}{GM_\odot^{2010}}
\newcommand{\Mn}{\mathcal M_\odot^{2010}}
\newcommand{\Tn}{\mathcal T_\odot^{2010}}
\newcommand{\Rn}{\mathcal R^\mathrm{N}_\odot}
\newcommand{\Ln}{\mathcal L^\mathrm{N}_\odot}

\newcommand{\kms}{km~s$^{-1}$ }
\newcommand{\ks}{km~s$^{-1}$}

\newcommand{\tef}{$T_{\rm eff}$ }

\newcommand{\ms}{$M_{\odot}$}
\newcommand{\rs}{$R_{\odot}$}

\begin{document}

\title{Call to adopt a nominal set of astrophysical parameters and
constants to improve the accuracy of fundamental physical properties of stars.
}


\author{Petr~Harmanec\altaffilmark{1}}
\affil{Astronomical Institute of the Charles University,
Faculty of Mathematics and Physics,\\
V~Hole\v{s}ovi\v{c}k\'ach 2, CZ-180 00 Praha 8, Czech Republic}
\email{hec@sirrah.troja.mff.cuni.cz}

\and

\author{Andrej Pr\v{s}a\altaffilmark{2}}
\affil{Villanova University, Department of~Astronomy \& Astrophysics,
800 Lancaster Ave, Villanova, PA 19085}

\shortauthors{Harmanec \&\ Pr\v{s}a}
\shorttitle{Obligatory astrophysical constants}

\date{Received March 17, 2011; accepted June 3, 2011 by PASP}

\begin{abstract}
The increasing precision of astronomical observations of stars and
stellar systems is gradually getting to a level where the use of slightly
different values of the solar mass, radius and luminosity, as well as
different values of fundamental physical constants, can lead to measurable
systematic differences in the determination of basic physical properties.
An equivalent issue with an inconsistent value of the speed of light was
resolved by adopting a nominal value that is constant and has no error
associated with it. Analogously, we suggest that the systematic error in
stellar parameters may be eliminated by: (1) replacing the solar radius
$R_\odot$ and luminosity $L_\odot$ by the nominal values that
are \emph{by definition} exact and expressed in SI units: $1 \Rnom = 6.95508 \times 10^8$\,m,
and $1 \Lnom = 3.846 \times 10^{26}$\,W; (2) computing stellar masses in terms of
$M_\odot$ by noting that the measurement error of the
product $GM_\odot$ is 5 orders of magnitude smaller than the error in $G$;
(3) computing stellar masses and temperatures in SI units by using the derived values
$\Mn = 1.988547 \times 10^{30}$\,kg and $\Tn = 5779.57$\,K; and (4) clearly stating the reference for
the values of the fundamental physical constants used. We discuss
the need and demonstrate the advantages of such a paradigm shift.
\end{abstract}




\section{INTRODUCTION AND MOTIVATION}

\begin{table}[!htb]
\caption[]{The values of 1 solar mass in the last 35 years. \label{table:MSun}}
\begin{flushleft}
\begin{tabular}{ll}
\noalign{\smallskip}\hline\hline\noalign{\smallskip}
\ms [$10^{30}$ kg]       & Source: \\
\noalign{\smallskip}\hline\noalign{\smallskip}
1.9891  & 1 (1976 IAU constants) \\
1.991 $\pm$ 0.002  & 2 (Handbook of Chem.~Phys.)\\
1.98892 $\pm$ 0.00025 & 3 (Astron. Almanac 1994) \\
1.9884 $\pm$ 0.0002 & 1 (Astron. Almanac 2011) \\
\noalign{\smallskip}\hline\noalign{\smallskip}
\end{tabular}
{\scriptsize {\bf Column ``Source" references:}\\
1... {\sl http://asa.usno.navy.mil/SecL/Comstants.html};
2... \citet{weast80};
3... Astronomical Almanac for the year 1994,
U.S. Government Printing Office, Washington and Her Majesty's Stationery
Office, London (1993)
}
\end{flushleft}
\end{table}

Notable improvements in the precision of astronomical observations
start to challenge our current models and understanding of the physical
processes in stars and stellar systems. Until recently, the accuracy of
fundamental stellar parameters derived from observations was limited by the stochastic
uncertainty of observations, but with new space-borne instruments
such as {\sl MOST} \citep{rucinski2003}, {\sl CoRoT} \citep{auvergne2009}
and {\sl Kepler} \citep{borucki2010}, systematic effects due to model
inadequacies are starting to play an important role.

One of the increasingly important deficiencies is the use of
the solar mass $M_\odot$, solar radius $R_\odot$, and solar luminosity
$L_\odot$ as units in which the fundamental stellar properties, deduced via
fitting the observations by appropriate models, are usually expressed.
Such stellar properties inevitably depend on the particular values of
the fundamental solar characteristics adopted by different researchers
in their studies. This inconsistency is rooted deeply in the literature
because the effect of discrepant parameters was often well within
the systematic and stochastic uncertainties of observations. However,
in certain fields, such as binary star research, the associated artifacts
become increasingly important, especially with the accumulation of precise
observations of systems with longer orbital periods.
They actually remain a fundamental issue across all fields because
eclipsing binaries (EBs) are the most accurate and commonly used
calibrators for the masses and radii of \emph{single} stars \citep{torres2010}.
A typical accuracy of stellar parameters for well-studied EBs is claimed
to be better than $\sim$2\% \citep{ander91}. The modern methods for
spectroscopic analysis, such as cross-correlation
\citep{simkin74, dacosta77, tonry79},
broadening functions \citep{rucin92,rucin98} and
disentangling \citep{simon94, korel1, korel2, ili2004},
as well as improved light-curve synthesis codes
\citep{wd71, wilson79, wilson94, wilson2008, prsa05, prsa06} can push
these limits even further. The systematic error due to inconsistently used
solar parameters hence propagates to the derived $M$-$R$-$L$-$T$ calibrations,
and has a notably adverse effect on the computed absolute scales and distances.
The situation, however, can be significantly improved if we are willing to part
with the \emph{current} values of canonical solar parameters and replace them
with the nominal values.

This issue is not novel in astronomy and a precedent has been set with the
exact value of the speed of light. In 1975, Resolution 2 of
the 15th Conf\'erence G\'en\'erale des Poids et Mesures (CGPM) proposed
the value of the speed of light in vacuum to be $c=299\,792\,458$\,m\,s$^{-1}$.
The value was chosen considering the excellent agreement to
$\delta c/c \sim 4 \times 10^{-9}$ among different measurement methods across all
wavelengths. In 1983, an argument was made that the \emph{unchanging}
speed of light needs to be maintained, notably for astronomy and geodesy,
and this was formalized by Resolution 1 of the 17th CGPM that defines
the speed of light to be the value recommended in 1975. Consequently,
1~m was {\sl redefined} to be the distance traveled by light in vacuum
in 1/299792458\,s. This way the nominal value of 1\,m depends on
a~fundamental natural constant rather than the other way around, as was
historically the case\footnote{See {\sl http://www.bipm.org/en/CGPM/db/17/1}}.

The re-definition of the bolometric magnitude followed at
the 23$^{\rm rd}$ General Assembly of the International Astronomical Union,
held in Kyoto in 1997 August, with adopting a resolution
\footnote{suggested by Dr. Roger Cayrel from Paris.} that specifies that the zero point of
the bolometric magnitude scale will no longer be defined through
the bolometric luminosity of the Sun, but rather by defining that
$M_{\rm bol} = 0.0$~mag corresponds to the bolometric luminosity
$L_{\rm bol} = 3.055 \times 10^{28} {\rm W}$. This introduces an
{\sl absolute scale} of bolometric magnitudes,
$M_{\rm bol} = 71.2125\,{\rm mag} - 2.5\log L$,
where the bolometric luminosity $L$ is given in watts. The convenience
of comparing $L$ to the solar value remains, since a new definition complies
with the most often quoted value of the solar bolometric magnitude:
$L_\odot = 3.846 \times 10^{26} \, {\rm W} \Rightarrow
M_{\rm bol,\odot} = 4.75$~mag.

\begin{table}[h]
\caption[]{The value of mean solar radius \rs\ adopted from various sources and
the corresponding solar effective temperature. The (almost negligible) errors
in values of the effective temperature are given solely by the error in the
determination of the Stefan-Boltzmann constant $\sigma$.}\label{r}
\begin{flushleft}
\begin{tabular}{cclcccc}
\noalign{\smallskip}\hline\hline\noalign{\smallskip}
\rs        & \tef       & Source\\
(km)      & (K)     &    \\
\noalign{\smallskip}\hline\noalign{\smallskip}
 696200.0&5776.702 $\pm$ 0.005 & A\\
 695990.0&5777.573 $\pm$ 0.005 & B\\
 695508.0&5779.575 $\pm$ 0.005 & C\\
 695835.3&5778.215 $\pm$ 0.005 & D\\
 695833.1&5778.224 $\pm$ 0.005 & E\\
 695770.0&5778.486 $\pm$ 0.005 & F\\
 695658.0&5778.952 $\pm$ 0.005 & G\\
\noalign{\smallskip}\hline\noalign{\smallskip}
\end{tabular}
\end{flushleft}
{\scriptsize
{\sl Abbreviations in column ``Source":} \hfill\break
A... IAU system of constants 1976;
B... Allen 3rd edition: \citet{allen3};
C... Allen 4th edition: \citet{allen4}, adopted from \citet{broda98};
D... this paper: a sinusoidal fit to all Cote d'Azur data 1975-1998;
E... this paper: a sinusoidal fit to good Cote d'Azur data 1975-1998;
F... \citet{tripa99};
G... \citet{habe2008}.}
\end{table}

Another important deficiency is the use of inconsistent (often outdated)
values of the fundamental physical constants without providing a reference
to the source of the value used. The derived constants cannot be made exact
since they \emph{observationally} depend on fundamental SI units.
The universal constant of gravitation $G$, for example, is defined as
the proportionality constant in Newton's law and is one of the most difficult
constants to measure to a high accuracy \citep{gillies1997}. The currently
recommended 2010 value of the constant by
the Committee on Data for Science and Technology (CODATA)\footnote{http://physics.nist.gov/cuu/Constants/index.html}
is $G = (6.67384 \pm 0.00080) \times 10^{-11}\,\mathrm{m}^3 \mathrm{kg}^{-1}\mathrm{s}^{-2}$.
 Since it is derived, this value is subject to change in the future
as more precise measurements become available. The solution to this problem is to commit to the use of constants
set forth by the IAU or CODATA and meticulously provide a reference to
the used value.

In this article we quantify the systematic effect of inconsistent
values of various parameters and propose to adopt the nominal values for
the mass, radius and luminosity of the Sun. We further propose to unify
the astrophysical constants across modeling tools and make a strong effort
to keep the values of the derived constants up to date with the IAU, CGPM
and CODATA resolutions. An excellent example is a recent review on accurate
stellar masses and radii by \citet{torres2010}, who quote values of all the
relevant constants used.

\section{QUANTIFICATION OF THE EFFECT}

The recently published Kepler Eclipsing Binary Catalog
\citep{prsa2011,slawson2011} contains over 2200 EBs in the 105~deg$^2$
field of view. The period distribution due to Kepler's uninterrupted baseline
does not suffer from any significant selection effects toward longer periods
($\sim$$10^2$ days) and a significant number of sources with
$P_\mathrm{orb} \geq 50$~d are emerging from the sample. Let us consider
as an~example a~binary system with two $1\,M_\odot$ components in
a~circular orbit with the orbital period of 200~d. The separation between
the components can be readily computed from Kepler's third law.
If we adopt the above quoted value of $G$
and compute the separation based on the $1\,M_\odot$ values listed in
Table \ref{table:MSun}, we will arrive at the following values
(in $10^9$\,m): $126.1624 \pm 0.0050$, $126.2026 \pm 0.0050$,
$126.1586 \pm 0.0050$, and $126.1476 \pm 0.0050$.
This accounts for
the relative error of $4.0 \times 10^{-5}$, or an absolute error of
$5.0 \times 10^6$\,m. A single event in Kepler long-cadence data
(30-minute exposure) can be timed to $\sim$6 minutes;
a~$P_\mathrm{orb} = 200$-day binary will have 18 events (nine primary and
nine secondary eclipses) over the 5 yr mission lifetime, which will reduce
the timing error to $\sim$1.4 minutes, or in relative terms to $5 \times 10^{-6}$.
This is an order of magnitude smaller than the effect of using inconsistent
solar-mass values, and will be the cause of significant systematics.
If short-cadence data (1-minute exposures) are available for the given target,
the timing is improved by another order of magnitude, and the systematics
will overpower the stochastic error.

The same is true for single stars as well. Consider a giant star with
a~radius of 30~\rs\ and an~equatorial rotational velocity of 5~\ks.
The period of rotation of such a star would then be 303.801~days for the
IAU~1976 value of the solar radius but 303.473~days for the recent
\citet{broda98} value (see~Table 2). Such a difference is readily
detectable after a few rotational periods covered by relevant observations.

Also worth considering is the dependence of the effective temperature on the
value of stellar radius through $L_\odot = 4\pi R_\odot^2 \sigma T_\mathrm{eff}^4$,
where $\sigma$ is the Stefan-Boltzmann constant. Table~\ref{r} lists several values of the solar radius,
either recommended in various compilations or derived from recent
measurements. For each value we give the corresponding effective temperature
calculated for the solar luminosity value of $L_\odot = 3.846 \times 10^{26}$\,W
and $\sigma = 5.670373(21)\times10^{-8}\,\mathrm{W\,m}^{-2}\mathrm{K}^{-4}$. The
systematic differences in $T_\mathrm{eff}$ values are larger than the propagated
errors.

\section{THE SOLAR UNITS}

\begin{deluxetable}{lcp{0.43\textheight}}
\tabletypesize{\scriptsize}
\rotate
\tablewidth{\textheight}
\tablecaption{Select examples of the commonly used equations utilizing the nominal
solar values proposed by this paper, and constants recommended by the IAU. \label{table:examples}}
\tablehead{
\colhead{Select example:} & \colhead{Equation:} & \colhead{Comments:}
}
\startdata
Kepler's third law: &
 $A [m] =  2927699260.629(74)P^{2\over{3}}(M_1+M_2)^{1\over{3}}$ &
$A$: semi-major axis in m, AU, and \Rnom; $M_1$ and $M_2$: point masses in
\Mnom; $P$: period in mean solar days.\\

& $A [AU] =  0.01957046077547(49)P^{2\over{3}}(M_1+M_2)^{1\over{3}}$ & \\
\\
& $A [\Rnom] =  4.20944009361(11)P^{2\over{3}}(M_1+M_2)^{1\over{3}}$ & \\
\\

Surface gravity: &
$\log g \ {\rm [cgs]} = 4.438307381330(33) + \log M - 2\log R$ &
$M$: stellar mass in \Mnom; $R$: stellar radius in \Rnom.
With the use of the nominal values of the solar mass and radius, the error only
depends on the error of the $G$\ms\ product constant and the conversion formula
becomes very accurate, highly exceeeding the current precision of our
knowledge of stellar masses and radii. \\
\\
Stellar bolometric magnitude: &
$M_{\rm bol} = 42.3689588(40) - 5\log R - 10\log T_{\rm eff} $ &
$R$: stellar radius in \Rnom; $T_\mathrm{eff}$: effective
temperature in K. The error in the conversion constant is due to the
uncertainty in the Stefan-Boltzmann constant $\sigma$. \\
\\
Parallactic radius: &
$R = 107.5457584245(22) \theta\,p^{-1}$ &
$R$: radius in \Rnom; $\theta$: angular stellar diameter in arcsec;
$p$: stellar parallax in arcsec. The constant is computed from the recommended
value of 1\,AU and
1\,pc by IAU 2009 ($149,597,870,700(3)$\,m and $3.085677581503(62)\times10^{16}$\,m,
respectively). \\
\\
Angular orbit diameter: &
$a = (215.0915168490(43))^{-1}A\,p$ &
$a$: angular orbit diameter in arcsec; $A$: semi-major axis in au;
$p$: stellar parallax in arcsec. The same
constants were used as above. \\
\\
Equatorial rotational velocity: &
$V = 50.57877\,R\,P_{\rm rot}^{-1}$ &
$V$: equatorial rotational velocity in \kms; $R$: stellar radius
in \Rnom; $P_\mathrm{rot}$: rotational period in mean solar days.
Note that the constant in
the above equation has no uncertainty thanks to our use on the nominal value
of the solar radius. \\

Mass functions: &
\parbox{0.45\textwidth}{
\begin{eqnarray}
 M_1\sin^3 i &=& 1.036149050206(78) \times 10^{-7} K_2 (K_1+K_2)^2 P(1-e^2)^{1.5} \nonumber \\
 M_2\sin^3 i &=& 1.036149050206(78) \times 10^{-7} K_1 (K_1+K_2)^2 P(1-e^2)^{1.5} \nonumber \\
      f_1(M) &=& 1.036149050206(78) \times 10^{-7} K_1^3 P(1-e^2)^{1.5} \nonumber \\
      f_2(M) &=& 1.036149050206(78) \times 10^{-7} K_2^3 P(1-e^2)^{1.5} \nonumber
\end{eqnarray}} &
$M_1$: stellar mass in \Mnom; $i$: inclination in deg; $K_{1,2}$: RV
semi-amplitudes in \mbox{\kms;} $P$: orbital period in mean solar days; $e$:
eccentricity. The uncertainty stems from the uncertainty in the
$G$\ms\ product. \\

Projected orbital sizes: &
\parbox{0.45\textwidth}{
\begin{eqnarray}
 a_1\sin i &=& 0.019771142 K_1P(1-e^2)^{0.5} \nonumber \\
 a_2\sin i &=& 0.019771142 K_2P(1-e^2)^{0.5} \nonumber \\
   A\sin i &=& 0.019771142 (K_1+K_2)P(1-e^2)^{0.5} \nonumber
\end{eqnarray}} &
$a_{1,2}$: distance of the orbiting component from the center of mass, in $\Rnom$;
$A = a_1+a_2$: semi-major axis in $\Rnom$; $i$: inclination in degrees; $K_{1,2}$:
RV semi-amplitudes in \mbox{$\mathrm{km}\,\mathrm s^{-1}$;} $P$: orbital period in days; $e$: orbital eccentricity.
Note that there is no error in the transformation constant. \\

Solar effective temperature: &
 ${\cal T}_{\rm eff.}^{\rm N} = 5779.5747(54)$ & ${\cal T}_{\rm eff.}^{\rm N}$
is the effective temparature of the Sun for the nominal value of the solar
radius\\
\enddata
\end{deluxetable}

It is customary in stellar research to express the basic physical properties
of the stars, such as their luminosity, mass or radius, in solar units.
While convenient, this is somewhat unfortunate, for the following reasons:
\begin{enumerate}
\item the values of the solar luminosity, mass and radius are subject
to continuous improvement thanks to the increasingly precise observational
techniques; and
\item solar luminosity and radius vary measurably with the solar cycle and perhaps also on other,
even shorter time scales \citep[see, e.g.,][]{sel2004}. Due to effects
such as mass loss via stellar wind or mass gain due to infalling material,
the change of the solar mass may become measurable on longer timescales.
\end{enumerate}

At the same time, some important quantities, such as the projected velocity
of the stellar rotation, are measured in absolute units
(\kms in this particular case) that depend directly on the adopted values of
the fundamental solar properties.

We suggest a simple solution to the problem of inconsistent use of various
different values of the solar units -- one that follows the precedent
for the speed of light and bolometric magnitude. Let us deprecate the use of
the current values of $R_\odot$ and $L_\odot$ with all their
uncertainties and time variations, and replace them
with the \emph{exact, nominal} units \Rnom\ and \Lnom, where their
numerical values are chosen to be close to the recently derived
(or adopted) values:
the solar radius from \citet{broda98}, also recommended in the 4$^{\rm th}$
edition of Allen's Astrophysical Quantities \citep{allen4}, and the solar luminosity
to comply with the above-mentioned IAU resolution on bolometric magnitude.

Once the values of $R_\odot$ and $L_\odot$ are made nominal, the relationship
between the nominal values $\Rn$ and $\Ln$, and the effective temperature $\Tn$ can be
determined readily from:
\begin{eqnarray}
   \Ln & = &  4\pi {\Rn}^2 \, \sigma \left(\Tn\right)^4, \mathrm{where} \nonumber\\
    \sigma &=& 5.670373(21)\times10^{-8}\,\mathrm{W\,m}^{-2}\mathrm{K}^{-4}. \nonumber
\end{eqnarray}
Note that the small error in the determination of the Stefan-Boltzmann
constant $\sigma$ propagates to negligibly small errors in the determination
of the effective temperature.

Unfortunately, the same cannot be proposed for the solar mass: while the universal
gravitational constant $G = (6.67384 \pm 0.00080) \times 10^{-11}\,\mathrm{m}^3 \mathrm{kg}^{-1}\mathrm{s}^{-2}$
is one of the least precisely determined fundamental constants in nature, the
product $GM_\odot = 1.32712442099(10)\times10^{20} {\rm m}^3{\rm s}^{-2}$ is
determined much more precisely \citep{petluz2010}. All physical parameters
that can be expressed in terms of $GM_\odot$ can thus be determined much more
accurately than the ones depending on $M_\odot$ alone. Table \ref{table:examples}
lists several imporant ones, most notably stellar masses in terms of solar mass,
and the absolute scale of the system in SI units. In particular, the mass
of a binary star component $j$ $(j=1,2)$ can be expressed as
$G M_j = [K_{3-j} P (K_1 + K_2)^2]/(2 \pi \sin^3 i)$, where $K_1$ and $K_2$ are
radial velocity semi-amplitudes in \ks, and $i$ is the inclination. If we divide
this expression by $GM_\odot$, we obtain $M_j/M_\odot$ in terms of $GM_\odot$,
which is much more precise than computing it with the more uncertain value of $G$.
However, both solar and stellar masses in SI units will still be limited by the
measurement error in $G$. If solar mass were to be made nominal, the precision in
$M_j/\Mn$ would be degraded by 5 orders of magnitude. To make the distinction clear,
we use the designation ${}^\mathrm{N}$ to represent nominal parameters, and ${}^{2010}$ to
denote parameters that are derived from the latest values of $GM_\odot$ and $\sigma$.

The proposed nominal and derived parameters to be used in computations are:

\begin{eqnarray}
1~\Rn  &=& 6.95508 \times 10^{8}~{\rm m} \label{rnom} \nonumber \\
1~\Ln  &=& 3.846 \times 10^{26}~{\rm W} \label{lnom} \nonumber \\
1~\GMn &=& 1.32712442099(10)\times10^{20} {\rm m}^3{\rm s}^{-2} \nonumber \\
1~\Mn  &=& 1.988547 \times 10^{30}~{\rm kg} \label{mnom} \nonumber \\
1~\Tn  &=& 5779.57\ \mathrm{K} \nonumber
\end{eqnarray}

\section {SELECT EXAMPLES}

There are a number of frequently used formulae where numerical constants are
affected by the adopted values of the solar mass, radius and luminosity.
We list some of them for the suggested nominal values in
Table \ref{table:examples}.
In all examples, the propagated errors of the fundamental physical constants
used are given in parentheses and denote the uncertainty of the last two
significant digits. The expressions for the semimajor axis $A$ from the
third Kepler law are given for the value of $A$ in km, AU, and
$\Rn$.

\section{CONCLUSIONS}
We presented the case for obsoleting the use of $R_\odot$, $L_\odot$ and $M_\odot$
as units and replacing two of them with their nominal counterparts $\Rn$ and $\Ln$
that are \emph{exact}. In all determinations of stellar masses expressed in \ms,
the product $\GMn$, known to a very high precision, should be used. For the
conversion of the thus derived stellar masses to SI units, $\Mn$ can be used.
Once the precision of the universal gravitational constant $G$ is improved, it
will be beneficial to deprecate the actual value of the solar mass and replace it
by an exact, nominal value $\mathcal M_\odot^\mathrm{N}$.

All the recommendations presented in this article would reduce the systematics that stem from using
discrepant values of fundamental properties of the Sun. This is a consequence of the true variations
of these values due to intrinsic effects such as the magnetic solar cycle, as well as
of a steady improvement in their determination via more precise observations.
We further implore the community to use fundamental physical constants recommended
by the IAU and/or CODATA and call for meticulous referencing of the used sources.

\begin{acknowledgements}
We appreciate several useful discussions on the subject, which we had with
M.~Bro\v{z}, P.~Mayer, D.~Vokrouhlick\'y, and P.~Zasche.
The research of PH was supported by the grant P209/10/0715 of
the Czech Science Foundation and
also from the Research Program MSM0021620860
{\sl Physical Study of Objects and Processes in the Solar System
and in Astrophysics} of the Ministry of Education of the Czech Republic. AP acknowledges the NASA/SETI subcontract 08-SC-1041.
We acknowledge the use of the electronic database from CDS Strasbourg and
electronic bibliography maintained by the NASA/ADS system.
\end{acknowledgements}

\bibliographystyle{aa}
\bibliography{citace}

\end{document}